\documentclass{desyproc}
\usepackage[colorlinks=true,linkcolor=black,urlcolor=black,citecolor=black]{hyperref}
\usepackage{sidecap}

\begin{document}
\title{Axions at the International AXion Observatory}

\author{{\slshape  Javier Redondo}\\[1ex]
University of Zaragoza, Zaragoza, Spain\\
Max Planck Institut f\"ur Physik, Munich, Germany}

\contribID{redondo\_javier}

\confID{11832}  
\desyproc{DESY-PROC-2015-02}
\acronym{Patras 2015} 
\doi  

\maketitle

\begin{abstract}
QCD axions with meV mass can be behind some stellar cooling anomalies and form all or part of the cold dark matter of the universe. We discuss on a proposed experiment to discover the solar flux of meV mass axions:  the International AXion Observatory (IAXO). 
\end{abstract}

\section{The meV mass axion frontier} 

The really low energy frontier of fundamental physics~\cite{Jaeckel:2010ni}  offers several pressing questions that have received much attention in the recent years~\cite{Baker:2013zta}. 
Very weakly-interacting sub-eV particles (WISPs) can arise as low energy manifestations of high energy completions of the standard model of particle physics and tend to be generically good dark matter candidates~\cite{Arias:2012az}. A central target is to discover the QCD axion, hypothetical particle predicted in the Peccei-Quinn mechanism to explain the puzzling absence of CP violation in the strong interactions, and only later realised as suitable for constituting the cold dark matter (CDM) and found to appear generically in string theories, prime candidates for describing quantum gravity. 
Actually, if the QCD axion exists, a sizeable amount of axion CDM is unavoidable (it turns out to be more natural for axion-like particles to solve the DM puzzle than to solve the strong CP problem). 

\begin{figure}[t!]
\centerline{\includegraphics[width=0.7\textwidth]{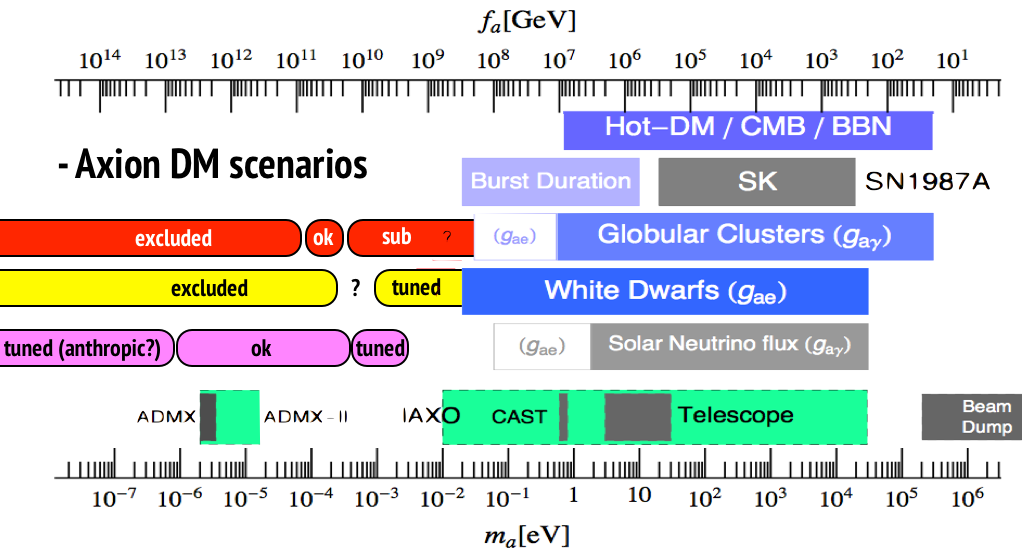}}
\vspace{-0.3cm}
\caption{\small Axion mass bands in the main CDM scenarios: \emph{after} (red for $N=1$, yellow for $N>1$) and \emph{before} (purple). Shown also are excluded bands from cosmology, stellar evolution and experiments together with sensitivities from ADMX-II and IAXO (green), from \cite{Essig:2013lka}. }\label{Fig:MV}
\label{figa1}
\end{figure}

The axion CDM density produced in the Big Bang depends on the axion mass, $m_a$, and the details of early cosmology. There are two basic scenarios: the axion field taking its initial conditions (typically after a phase transition) either \emph{after} cosmic inflation or \emph{before}. 
In the \emph{after} scenario, the initial conditions are random in causally disconnected regions and a network of global strings forms through the Kibble mechanism.  
QCD instantons generate a potential for the axion with a set of $N$ CP-conserving minima ($N$ depends on the UV completion of the axion model), which is strongly suppressed at high temperatures but becomes relevant close to the color confinement phase transition, $T_{\rm QCD}$.  
By then, the field relaxes to one minimum and oscillates around it with its amplitude damped by the universe expansion. The harmonically oscillating field is a coherent state of very non-relativist quanta (axions), a cold dark matter fluid. The network of strings and domain walls developed around $T_{\rm QCD}$ is unstable if $N=1$ and decays into a second population of CDM axions. The first contribution is computable but the second (which seems to dominate) has to be extrapolated over many orders of magnitude from numerical simulations. The latest simulations~\cite{Kawasaki:2014sqa} give\footnote{A recent analysis~\cite{Fleury:2015aca} challenges the interpretation of the simulations, adding to a longstanding controversy.},  
\begin{equation}
\Omega_{a} h^2 = 0.12\left(\frac{1}{3.4}+\frac{2.4\pm 1.12}{3.4}\right)\left(\frac{108\, \rm \mu eV}{m_a}\right)^{1.187}
\quad ; \quad ({\rm misalignment}\,+\,{\rm strings})
\end{equation}
suggesting an axion CDM mass $m_a=105\pm25\, \mu$eV labelled `ok' in Fig.~\ref{figa1} in red. 
Smaller values overproduce DM and are excluded, and larger ones imply a subdominant fraction of the CDM $\gtrsim (0.11/20)^{1.187}=0.2\%$ for $m_a\lesssim 20$ meV.   
If $N>1$ the string-wall network is stable and thus ruled out unless a small energy breaks the degeneracy of vacua. This breaking needs to be extremely small because it tends to displace the minimum away from CP conserving and spoil the solution of the strong CP problem. The hecatomb of strings and walls gets delayed by the small degeneracy breaking, making CDM axions less diluted and more abundant today, favouring much larger axion CDM masses (see yellow labelled `tuned' in Fig.~\ref{figa1}).  
Finally, in the \emph{before} scenario, inflation makes homogeneous the axion field in our observable universe and dilutes away strings and walls. The observed amount of CDM can be obtained for any $m_a<$ meV by invoking the appropriate axion initial condition. Excluding fine tunings of $10\%$ to the bottom or top of the potential the preferred range is  1 $\mu$eV$<m_a<0.5$ meV (purple band labelled `ok'). 

\begin{SCfigure}
\includegraphics[width=0.6\textwidth]{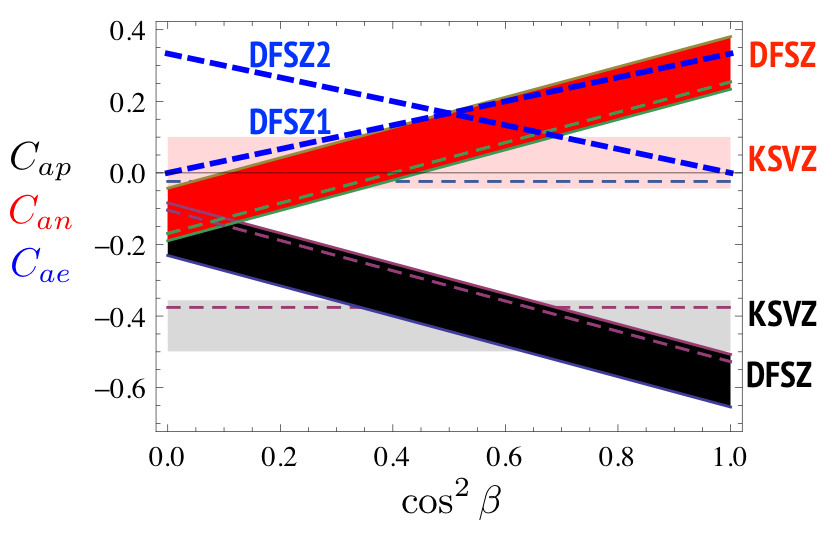}\vspace{-0.4cm} \caption{\small Axion-proton, neutron and electron couplings (black, red, and blue resp.) with low-energy-QCD error bars ($m_u/m_d=0.56^{+0.04}_{-0.26}$) in the KSVZ and DFSZ 1 and 2 models. Here $C_{af}$ is the coupling normalised with the axion decay constant $f_a$ and the fermion mass $m_f$: $C_{af}=g_{af}\times f_a/m_f$ ($\tan\beta$ is the ratio of extra Higgs fields in the model~\cite{Dias:2014osa}).}\label{Fig:MV}
\label{figa2}
\end{SCfigure}

Figure \ref{figa1} shows the CDM regions together with the exclusion bounds from cosmology astrophysics and experiments and makes a very clear point: would DM be made of $\sim$meV mass axions,  we shall then expect some effects in astrophysics too. 
Indeed, for several years now, there have been increasing claims of anomalies in the cooling of certain types of stars that could be attributable to QCD axions. We shall here briefly name them and show that a consistent axion model exists which fits every claim, constituting a prime target for a next generation helioscope: IAXO. 
The axion Lagrangian defines conventions for the axion coupling to photons and fermions,  
\begin{equation}
{\cal L}_a = \frac{1}{2}(\partial_\mu a)(\partial^\mu a)-\frac{1}{2}m_a^2a^2 
- \frac{g_{a\gamma}}{4}F_{\mu\nu}\tilde F^{\mu\nu} a
+\sum_f\frac{g_{af}}{2}\bar \Psi_f \gamma^\mu\gamma_5 \Psi_f \partial_\mu a . 
\end{equation}
In UV-complete axion models the couplings are related through a few parameters. Here we use KSVZ and DSFZ (with variants 1 and 2) models as exposed in Ref.~\cite{Dias:2014osa} and show the electron, proton and neutron couplings in Fig.~\ref{figa2}.  

White dwarfs (WDs) are degenerate stars not massive enough to fuse C and O into heavier nuclei, which just cool down by neutrino thermal emission from the core and surface electromagnetic radiation. The WD luminosity function (number of WDs per unit luminosity) decreases if there is an extra channel for plasma energy loss~\cite{Raffelt:1985nj}. Recently compiled luminosity functions tend to fit better expectations if the emission of axions in nucleus-electron bremsstrahlung is added\footnote{The anomalous cooling of the variables G117-B15A and R548 could also be due to axion emission, but the interpretation depends on the proper identification of the oscillating mode, so we have not included it in this analysis.
} with a coupling strength~\cite{Bertolami:2014wua}, 
\begin{equation} 
\label{WD}
g_{ae} =(1.4\pm 1.4)\times 10^{-13}. 
\end{equation}

Axion emission from the red giant star cores cools the plasma delaying the Helium flash, which happens at a larger core mass and thus becomes brighter. The study of the red giant branch of M5~\cite{Viaux:2013hca} yields a 95\% CL upper bound $g_{ae}<4.3\times 10^{-13}$ but a $1$-$\sigma$ preferred region, 
\begin{equation} 
\label{RG}
g_{ae} =(2\pm 1.5)\times 10^{-13}.  
\end{equation}

The swift cooling of the neutron star CAS A observed for 10 years by CHANDRA seems to confirm neutrino pair emission in neutron Cooper pair formation, $n+n\to {}^3{\rm P}_2+\bar \nu\nu$, as the responsible cooling mechanism but theoretical emission rates fall short by a factor of two~\cite{Leinson:2014ioa}, accountable among others~\cite{Leinson:2014cja} by a similar axion emission process, $n+n\to {}^3{\rm P}_2+ a$ if the axion-neutron coupling were~\cite{Leinson:2014ioa}, 
\begin{equation} 
\label{NS}
g_{an} =(3.8\pm 3)\times 10^{-13}.   
\end{equation}

Other interesting anomalies have been presented in this workshop~\cite{anomalies} and elsewhere~\cite{anomalies2}. A full analysis is in progress and shall be reported elsewhere~\cite{maurizio}.  We advance that some of them are not quantitative enough and some others cannot be directly attributable to QCD axions because of the strong constraint on the axion-proton coupling derived from the duration of the detected neutrino pulse from SN1987a~\cite{Raffelt:2006cw},  
\begin{equation} 
\label{SN}
g_{ap} < 8\times 10^{-10} , 
\end{equation} 
certainly in need of refinement from new simulations and data from a next galactic SN. 
\begin{figure}[h]
\centerline{\includegraphics[width=0.99\textwidth]{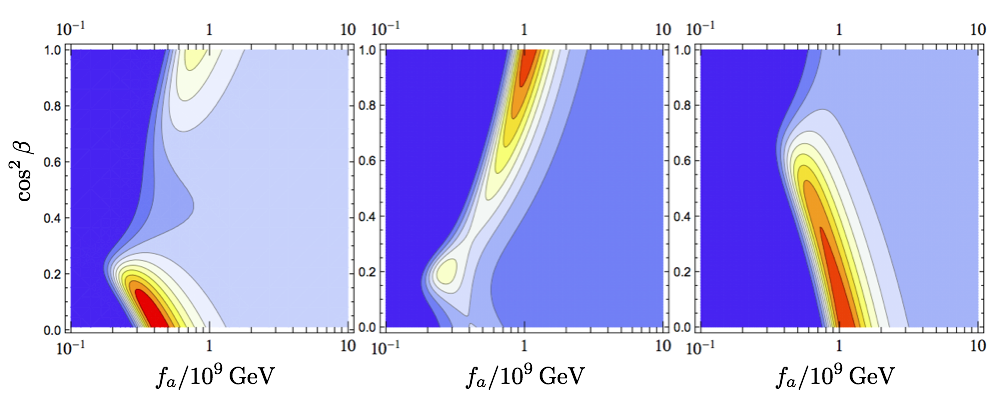}}
\vspace{-0.3cm}
\caption{\small Isocontours of relative $\chi^2$ (blue to red from max to min) for DFSZ models as function of the axion decay constant $f_a$ and $\cos^2\beta$ ($\tan\beta$ is the ratio of extra Higgs fields in the model~\cite{Dias:2014osa}).}
\label{figa3}
\end{figure}

We can now use a $\chi^2$ function of the different exotic energy losses which led to constraints \eqref{WD}-\eqref{SN} to estimate the microscopic parameters of the axion models.  
The KSVZ model has $C_{ae}\simeq 0$ so in principle it could only fit the NS cooling anomaly~\eqref{NS} while respecting the SN constraint \eqref{SN} but this is not the case because $g_{an}$ is larger than $g_{ap}$, see Fig.~\ref{figa2}. 
For DFSZ models, the $\chi^2$ including only NS and SN data (Fig.~\ref{figa3} left) is maximal at two points, which correspond to large neutron to proton coupling happening at $\cos^2\beta=1,0$ for large $f_a$ ($\sim 10^9\, {\rm GeV}$) and small $f_a$ solutions ($\sim 3\times 10^8\, {\rm GeV}$). 
Including WD and RG data in DFSZ1 (Fig.~\ref{figa3} center) we see all fitting in both the small and large $f_a$ points, with a larger tension in the small $f_a$. 
The DFSZ2 scenario (Fig~\ref{figa3} right) fits also the WD and RG anomalies respecting the SN constraint but cannot fit the NS simultaneously. In summary, there are two interesting targets 
\begin{eqnarray}
\nonumber
{\rm P1}: &   f_a\sim 10^9\, \rm GeV &\quad {\rm explains} \quad\rm RG+WD+NS\, (DFSZ1),\, RG+WD\, (DFSZ2),\\
\nonumber
{\rm P2}: &   f_a\sim 3\times 10^8\, \rm GeV &\quad {\rm explains} \quad\rm RG+WD+NS\, (DFSZ1),
\end{eqnarray}
which correspond to masses $m_a\simeq 0.6$ and 20 meV resp. 

\section{IAXO: International AXion Observatory}
Searching for meV mass axions seems to be within the reach of a future helioscope~\cite{Irastorza:2011gs}, or perhaps a future generation of 5th force searches~\cite{Arvanitaki:2014dfa} but the direct detection of meV DM axions seems extremely challenging. 
The helioscope technique~\cite{Sikivie:1983ip} aims at detecting the copious flux of axions produced in the solar core via either Primakoff process ($\propto g_{a\gamma}^2$) or the ABC processes ($\propto g_{ae}^2$)~\cite{Redondo:2013wwa}. 
Solar axions of energy $\omega$ convert coherently into detectable X-rays along an homogeneous transverse magnetic field $B$ of length $L$ with a probability 
\begin{equation}
P(a\to \gamma) \simeq \frac{(2g_{a\gamma} B \omega)^2}{m_a^4}\sin^2\left(\frac{m_a^2L}{4\omega}\right) . 
\end{equation}   
The most successful helioscope to date, CAST, uses a $9$\,T, $10$\,m long LHC decommissioned magnet mounted on a movable platform to track the Sun for $\sim 2$\,h/day with CCD and micromegas detectors at its bore ends.  It mostly suffers from a small bore aperture ($\sim 14.5$\,cm$^2$) and issues to track the Sun far outside horizontal positions, consequences of the dipole being designed as part of a proton collider unaware of its today's axionic duties. 

Members of the CAST collaboration are seeding a collaboration to build the first right-to-scale axion helioscope: the International AXion Observatory~\cite{IAXO}. The central target is to build a new magnet dedicated solely to axion physics not to suffer from any of the constraints inherited from a recycling experiment. In~\cite{Irastorza:2011gs}, a preliminary study based on the CAST experience demonstrated that the technologies matured in CAST would allow for an improvement of up to 6 orders of magnitude in signal/noise beyond CAST with the use of a new toroidal magnet operated with X-ray focusing optics and state of the art Micromegas. Since then, the collaboration has grown to ${\cal O}(100)$ scientists from ${\cal O}(40)$ institutions, a conceptual design report was produced~\cite{Armengaud:2014gea} and a LOI presented at CERN~\cite{LOI} (although the site of IAXO is by no means yet decided). A 20-m long, 5-m diameter toroid to be operated in a fully steerable platform has been designed~\cite{Shilon:2012te} in collaboration with the CERN magnet labs. It would have 8 warm bores of 0.6\,m diameter with an average field of 2.5\,T (5\,T peak field). The X-ray optics to be mounted at the bore's ends has been designed~\cite{optics} by the IAXO groups at LLNL, Columbia U. and DTU Denmark.  Micromegas detectors have been shown levels of $8\times 10^{-7} $counts/(keV\,cm$^2$\,s) in the CAST 2014 run and $10^{-7} $counts/(keV\,cm$^2$\,s) in a dedicated prototype at the Canfranc underground lab~\cite{garcia}, which advance the ambitious goal $10^{-8} $counts/(keV\,cm$^2$\,s) as realistic. 
New groups in IAXO have brought expertise in other detection technologies such as Gridpix/InGrid,  MMCs and low-noise CCDs, presented also in this workshop.  

\begin{figure}[t!]
\centerline{\includegraphics[width=\textwidth]{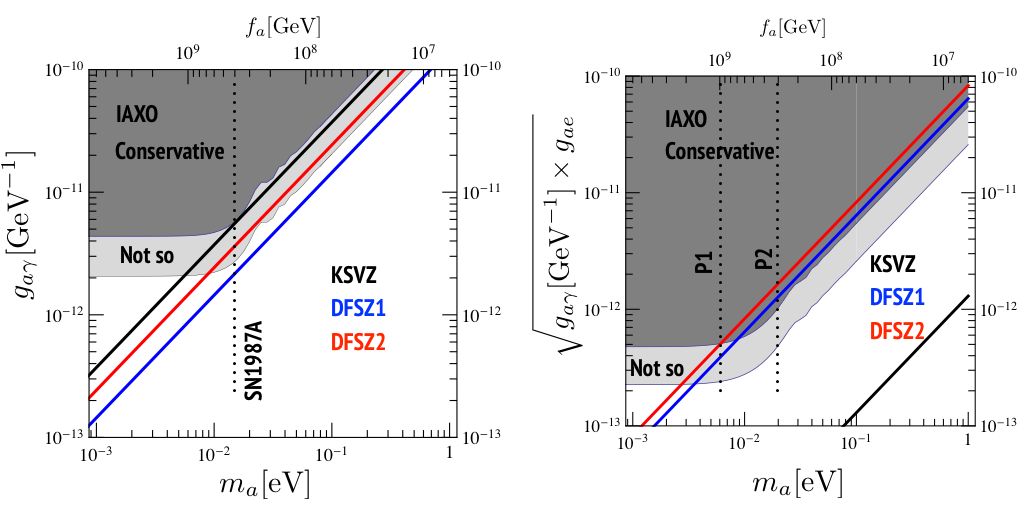}}
\vspace{-0.3cm}\caption{
\small IAXO sensitivity for Primakof (left) and ABC (right) solar axions in the ``conservative'' and ``not so'' configurations. Axion models KSVZ, DFSZ1 and 2 as black, blue and red lines and the target points P1 and P2 motivated by astro-hints. }\label{Fig:MV}
\label{figa4}
\end{figure}

In Fig.~\ref{figa4} we show in dark gray the sensitivity of a 3-year data campaign of IAXO to solar axions of Primakoff (left) and ABC (right) origin, with the parameters shown to be Conservative in the CDRs~\cite{Armengaud:2014gea,Shilon:2012te} but a background figure of $10^{-8} $counts/(keV\,cm$^2$\,s). 
The ABC flux of target P2 is in the discoverable region and of P1 only in the DFSZ2 model. We are considering improvements over the base design to raise the signal to noise up to a maximal factor of $\sim 20$ (``Not so'' lightgray region), which would cover all the interesting points at high confidence and consistently improve over the SN1987a constraint scanning unconstrained parameter space where, as we argued before, axions can constitute  all or part of the CDM of the universe.   

IAXO has a large potential impact beyond discovering solar QCD axions. The flux of axions from the core collapse of Betelgueuse could be detected if IAXO is pointed at it with an early warning~\cite{Raffelt:2011ft}.  Other WISPs from the Sun could be detected, such as axion-like particles or hidden-photons~\cite{Redondo:2008aa}. 
Axion-like particles with a photon-coupling $\sim 10^{-11}$GeV$^{-1}$ have been invoked as a solution for the anomalous transparency of the universe to high-energy photons~\cite{anomalies} and will be either found or excluded by IAXO. Finally, we are studying the possibility of hosting direct axion CDM experiments in the IAXO magnet~\cite{IAXODM}.

\begin{footnotesize}

\end{footnotesize}


\end{document}